\documentclass[twocolumn]{openjournal}
\usepackage{amsmath,amssymb,physics,bm,esint}
\usepackage{commath}
\usepackage{orcidlink}
\usepackage{color}
\usepackage{listings}
\usepackage{graphicx}
\usepackage{textcomp}
\usepackage{soul}
\usepackage{comment}
\usepackage{placeins} % for \FloatBarrier
\usepackage{xlmacros}
\usepackage{tikz}
\usepackage{lineno}

\DeclareRobustCommand{\VAN}[3]{#2}
\let\VANthebibliography\thebibliography
\def\thebibliography{\DeclareRobustCommand{\VAN}[3]{##3}\VANthebibliography}

\newcommand{\RAIL}{\texttt{RAIL}}
\newcommand{\FlexZBoost}{\texttt{FlexZBoost}}
\newcommand{\BPZ}{\texttt{BPZ}}

\newcommand{\pz}{photo-$z$}

\newcommand{\sigmaNMAD}{\sigma_{\rm NMAD}}

\begin{document}

\title{Weak-Lensing Shear Response for Photometric Redshift-Based Tomographic Binning}

\author{Xiangchong Li}
\email{xli6@bnl.gov}
\affiliation{Brookhaven National Laboratory, Bldg 510, Upton, New York 11973,
USA}

\author{Tianqing Zhang}
\affiliation{Department of Physics and Astronomy and PITT PACC, University of Pittsburgh, Pittsburgh, PA 15260, USA}
\affiliation{Kavli Institute of Particle Astrophysics and Cosmology, Stanford, CA, 94305, USA}
\affiliation{SLAC National Accelerator Laboratory, Menlo Park, California, 94025, USA}

\author{Rachel Mandelbaum}
\affiliation{Department of Physics, McWilliams Center for Cosmology and Astrophysics, Carnegie
Mellon University, Pittsburgh, PA 15213, USA}

\author{the LSST Dark Energy Science Collaboration}

\submitted{Received Month ??, 2025; accepted Month ??, 2025}

\begin{abstract}
Dividing source galaxies into tomographic redshift bins is a cornerstone of
modern weak gravitational lensing analyses, enabling measurements of the
growth of cosmic structure and the nature of dark energy. In practice,
these tomographic bins are defined using photometric redshift (\pz{})
estimates. However, correlations between \pz{} estimates and weak lensing
shear can introduce redshift-dependent selection biases in the measured
shear signal; if left uncorrected, these biases distort the inferred
amplitude and redshift evolution of the lensing signal, and in turn bias
the measurement of the growth of cosmic structure across cosmic time. In
this paper, we extend the analytical self-calibration for shear measurement
(\anacal{}) framework to account for \pz{}-based selection biases in
tomographic weak lensing analyses by propagating shear responses through
the selection process. This approach eliminates the need for external image
simulations to calibrate this correction. As a first sanity check on real
data, we validate the \pz{} estimates derived from \anacal{} fluxes on the
Rubin Observatory Data Preview 1 dataset, and find that they reach
\pz{} quality comparable to, and at high redshift slightly better than,
the standard LSST estimates. We then validate the shear
calibration on LSST-like image simulations with blending at the expected
LSST Y10 depth, with two representative \pz{} algorithms---a template-fitting method and a
machine-learning method---and show that the multiplicative shear bias
induced by \pz{} selection remains within the LSST ten-year requirement
$|m| < 3\times 10^{-3}$ across all five tomographic bins for both
algorithms. These results establish \anacal{} as a self-consistent
pipeline for tomographic weak lensing science in upcoming LSST analyses.
\end{abstract}

\keywords{
    gravitational lensing: weak --- galaxies: distances and redshifts ---
    cosmology: observations --- techniques: image processing
}

\maketitle

\section{INTRODUCTION}
\label{sec:intro}

Large-scale photometric imaging surveys currently under construction or in
operation are opening a new window on precision cosmology \citep{DarkEnergy}.
A principal science goal of these programs is to constrain the standard
cosmological model and to elucidate the physical origin of cosmic acceleration.
Weak gravitational lensing---the coherent distortion of distant galaxy images
by intervening large-scale structure---offers one of the most direct probes of
the matter distribution and its growth over time
\citep{rev_wl_Bartelmann01, rev_cosmicShear_Kilbinger15,
rev_wlsys_Mandelbaum2017, rev_wl_PratBacon2026}. Over the past decade, the
ongoing ``Stage III'' photometric surveys have demonstrated the cosmological
power of this probe through tomographic cosmic-shear and ``3$\times$2pt''
analyses, including the Dark Energy Survey \citep{desy3, desy6}, the
Kilo-Degree Survey \citep{kids1000_Li2023, kidsLegacy_Wright2025, kidsLegacy_Stolzner2025},
and the Hyper Suprime-Cam Subaru Strategic Program
\citep{hscy3, hscy3_peaks, hscy3_3x2pt_Zhang2026a, hscy3_3x2pt_Zhang2026b},
all of which derive their cosmological constraints from tomographic
measurements of weak lensing shear binned by photometric redshift.

Three flagship ``Stage IV'' programs will together image billions of galaxies
over the next decade: the Vera C.\ Rubin Observatory Legacy Survey of Space
and Time (LSST, \citealt{LSSTOverviwe2019}), the European Space Agency's
\textit{Euclid} mission \citep{Euclid2011}, and NASA's Nancy Grace Roman
Space Telescope \citep{Roman2020}. Their shared science requirements impose
tight bounds on multiplicative shear bias, typically at the $\sim 10^{-3}$
level \citep{WLsys_Massey2013, LSSTRequirement2018}, motivating the
development of ever-more rigorous calibration methods.

At the level of a single source, weak lensing distorts the observed galaxy
image through two effects: an isotropic magnification described by the
convergence $\kappa$, which rescales the image size and flux, and an
anisotropic stretching described by the two-component shear
$(\gamma_1,\gamma_2)$, which transforms the apparent galaxy shape. Because
intrinsic galaxy shapes are randomly oriented, the ensemble-averaged
ellipticity is an unbiased shear estimator---provided the shear response of
each shape measurement is accurately characterised \citep{metacal_Huff2017}.
Constructing a well-calibrated response factor is therefore the central
challenge of any weak-lensing shear analysis.

Several methods now calibrate shear without relying on external image
simulations, at least when blended sources lie at the same redshift.
\metadet{} \citep{metaDet_Sheldon2020, metaDet_LSST2023} achieves this
through numerical self-calibration of both detection and selection effects.
\anacal{} \citep{Anacal_Li2023, Anacal_Li2025} takes an orthogonal approach,
computing the full shear-response chain analytically from smoothed pixel
values upward; consequently, it processes each galaxy in under a millisecond,
orders of magnitude faster than simulation-based alternatives.
\BFD{} \citep{BFD_Bernstein2014, BFD_Bernstein2016} frames shear inference
as a Bayesian problem, but may still require percent-level simulation inputs
to correct for detection bias in blended fields.
Of these methods, \metadet{} can also correct the selection
bias from tomographic bin assignment without external image simulations, by
repeating the tomographic binning on the catalogues measured from its
(typically five) sheared images; \anacal{} achieves the same correction
analytically at the catalogue level, without shearing images or repeating
detection and measurement, and is therefore the simplest and most
computationally efficient approach.

Because \anacal{} is the framework we build on in this work,
we now describe it in more detail. \citet{Anacal_Li2023} and
\citet{Anacal_Li2024} showed that the four dominant
sources of shear bias---detection bias, selection bias, galaxy model bias,
and PSF shape leakage---can be addressed within a single pixel-level framework
accurate to second order in shear. The approach begins with the analytically
derived shear response of each Gaussian-smoothed pixel and propagates it to
any downstream observable via the chain rule, without perturbing images
numerically. Building on this foundation, \citet{Anacal_Li2025} extended the
framework to include an analytical correction for noise bias in shear
estimation, implemented through the renoising procedure described
in Section~\ref{subsec:anacal_review}.

Tomographic weak lensing is the primary analysis framework for extracting
cosmological information from shear catalogues. In this approach, source
galaxies are divided into photometric-redshift bins, and shear correlations are
measured within and across bins \citep{rev_cosmicShear_Kilbinger15}. However,
photometric redshift (\pz{}) estimates depend on galaxy observables such as
flux, size, and morphology, which are themselves perturbed by lensing shear.
As a result, tomographic bin assignment becomes correlated with shear,
introducing a shear-dependent selection effect (tomographic selection
bias). Because this bias depends on redshift, it propagates differently
into each tomographic bin and, if left uncorrected, distorts the inferred
amplitude and redshift evolution of the lensing signal. In turn, it biases
the measurement of the growth of cosmic structure across cosmic time,
which is one of the key cosmological observables targeted by tomographic
weak lensing analyses.

In this paper, we extend the \anacal{} formalism to correct shear biases
introduced by \pz{}-based tomographic selection. By analytically propagating
the shear response through the tomographic selection step, we correct the
selection bias induced by redshift binning without relying on external image
simulations for this calibration. We first verify, on the Rubin Observatory
Data Preview 1 (DP1) dataset \citep{dp1_dataset, dp1_paper}, that \pz{} estimates derived
from \anacal{} fluxes are of comparable quality to those derived from the
standard LSST Science Pipelines photometry \citep{bosch2018}, establishing the feasibility of estimating
\pz{} from the same flux measurements that feed the \anacal{} shear
estimator. We then validate the analytical selection-bias correction on
LSST-like image simulations with blending at the expected LSST Y10 depth,
using \anacal{}
to measure galaxy fluxes together with their shear responses and estimating
\pz{} with the \RAIL{} framework \citep{RAIL} for two representative
algorithms, \BPZ{} (template-fitting) and \FlexZBoost{} (machine-learning).
We note that external simulations may still be
required to calibrate the source redshift distribution in the presence of
blending between galaxies at different redshifts, which experience different
shears \citep{DESY3_BlendshearCalib_MacCrann2021}. In this work, however, we
focus specifically on the shear-estimation bias induced by \pz{}-based
selection and leave the treatment of multi-redshift blending to future work.

We emphasise that this correction requires careful testing of
the full \pz{} pipeline before it is applied to data: we validate the
\anacal{} selection-bias correction with image simulations for each \pz{}
estimation code, as we do here for \BPZ{} and \FlexZBoost{}. The image
simulations serve only to validate the pipeline; the shear calibration itself
does not depend on them.

This paper is organised as follows:
Section~\ref{sec:method} describes the \anacal{} framework, the \pz{}
algorithms, and the \anacal{} correction for \pz{}-based selection bias.
Section~\ref{sec:dp1} validates the performance of \pz{} measurements based
on \anacal{} fluxes using the DP1 dataset.
Section~\ref{sec:validation} presents the validation of the selection bias
correction on LSST-like image simulations.
Section~\ref{sec:summary} provides a summary and outlook.
Together, Section~\ref{sec:dp1} and
Section~\ref{sec:validation} establish that \anacal{} fluxes can be used for
\pz{}-based tomographic binning in LSST: Section~\ref{sec:dp1} demonstrates
their \emph{precision and efficiency}, showing that they reach a \pz{} quality
comparable to the photometry measured by the LSST Science Pipelines and can
therefore be used for tomographic binning on real observations, while
Section~\ref{sec:validation} demonstrates the \emph{accuracy} of the resulting
calibration, showing that the residual shear selection bias from tomographic
binning meets the LSST DESC requirement.

\section{SELECTION BIAS CORRECTION}
\label{sec:method}

In this section we present the \anacal{} framework for analytically correcting
the shear selection bias introduced by \pz{}-based tomographic binning. We
first introduce the two \pz{} algorithms used in this work---\BPZ{} and
\FlexZBoost{}---both accessed through the \RAIL{} framework
(Section~\ref{subsec:rail}). We then review the analytical pixel-level shear
response and the \FPFS{} shear estimator (Section~\ref{subsec:anacal_review}).
Finally, Section~\ref{subsec:pz_selection} describes how the \pz{}-based
selection bias is corrected within the \anacal{} formalism.

\subsection{Photo-$z$ Estimates with \RAIL{}}
\label{subsec:rail}

We use \pz{} estimates produced with the Redshift Assessment Infrastructure
Layers \citep[\RAIL{};][]{RAIL}, an open-source, modular framework that
provides a unified interface for training, applying, and evaluating \pz{}
algorithms for large photometric surveys. \RAIL{} has been validated on the
DP1 dataset \citep{dp1-pz}, which provides
the \pz{} measurements adopted in this work. Below we summarise the two
algorithms used, together with the training data and performance metrics
most relevant to our analysis, and refer readers to \citet{dp1-pz} for full
details.

We employ two \RAIL{}-supported algorithms, one representative of each of
the two main classes of \pz{} estimators: \BPZ{} for template fitting and
\FlexZBoost{} for machine learning.

\emph{Template-fitting} algorithms estimate redshifts by comparing observed
galaxy photometry to a library of spectral energy distribution (SED) templates
spanning a range of galaxy types and redshifts \citep{dp1-pz}. For each
template at each trial redshift, synthetic fluxes are computed by redshifting
the SED and convolving with the survey filter responses; $\chi^2$-based
likelihoods are combined with Bayesian priors as a function of magnitude and
galaxy type to produce a one-dimensional redshift posterior PDF. We use \BPZ{}
\citep{Benitez:2000, Coe:06}, which employs the SED library of \citet{Ilbert:09}
consisting of elliptical, spiral, and starburst templates, with a grid of
internal dust extinction values applied. Template-fitting algorithms can
extrapolate to redshifts not represented in the training data, making them
potentially more robust at high redshifts, but their accuracy is limited by the
fidelity of the assumed SED models to the real galaxy population.

\emph{Machine-learning} algorithms predict redshifts by learning empirical
mappings from multiband photometry to redshift from a labelled training set.
They can be less susceptible to SED-model mismatch than template-fitting methods,
but require a training sample representative of the full galaxy population. We
use \FlexZBoost{} \citep{Izbicki:2017}, which employs boosted decision trees to
model the non-linear colour--redshift relation and outputs a full posterior PDF
over redshift.

Both algorithms output a full one-dimensional posterior PDF $p(z)$ for
every galaxy. Following \citet{dp1-pz}, we adopt the mode of the
posterior, $z_\mathrm{mode} \equiv \mathrm{argmax}_z\, p(z)$, as our
point-estimate redshift throughout this work. The mode is a simple,
standard summary of the PDF and is the natural choice for hard
tomographic cuts; we note that any single point estimate (mode, mean,
or median) discards the multimodality characteristic of catastrophic
outliers \citep{dp1-pz}, motivating the use of the full PDF for bin
assignment as a future extension (Section~\ref{sec:summary}).

We quantify \pz{} performance using the normalised deviation
\begin{equation}
\label{eq:deltaz}
    \Delta z = \frac{z_\mathrm{mode} - z_\mathrm{ref}}{1 + z_\mathrm{ref}}\,,
\end{equation}
where $z_\mathrm{ref}$ is the reference redshift. For a given sample of $N$
galaxies, the mean bias is
\begin{equation}
\label{eq:Edeltaz}
    \mathbb{E}[\Delta z] = \frac{1}{N} \sum_{i=1}^{N} \Delta z_i\,,
\end{equation}
the scatter is measured by the normalised median absolute deviation (NMAD)
\begin{equation}
\label{eq:sigmaNMAD}
    \sigmaNMAD = 1.48 \times \mathrm{median}\!
    \left(|\Delta z - \mathrm{median}(\Delta z)|\right)\,,
\end{equation}
and the catastrophic outlier rate $\eta_{0.15}$ is the fraction of
galaxies with $|\Delta z| > 0.15$. Performance metrics for both algorithms
on the DP1 test set are presented in Section~\ref{sec:dp1}.

\subsection{Review of the Analytical Calibration Framework}
\label{subsec:anacal_review}

We briefly review the \anacal{} framework
\citep{Anacal_Li2023, Anacal_Li2024, Anacal_Li2025} and refer readers to
those papers for detailed derivations. The key idea is to compute the shear
response of each smoothed image pixel analytically, and then propagate those
responses to downstream observables---including fluxes, detection statistics,
selection variables, and shape estimators---through the chain rule.

\subsubsection{PSF Correction and Noise Bias Removal}

Let $f_p(\vx)$ denote the observed image after convolution with the
point spread function (PSF) $p(\vx)$. Because the PSF is generally
anisotropic and spatially varying,
\anacal{} removes its imprint in two steps: deconvolution of $p(\vx)$ in
Fourier space, followed by reconvolution with an isotropic Gaussian kernel
of width $\sigma_h$,
\begin{equation}
    h(\vk) = \exp\!\left(-\frac{k^2 \sigma_h^2}{2} \right)\,.
\end{equation}
Rounding the effective PSF in this way is a prerequisite for the
closed-form pixel shear response derived in \citet{Anacal_Li2023}.

To suppress stochastic noise bias, \anacal{} adopts the renoising
procedure of \citet{Anacal_Li2025}, analogous to metacalibration
\citep{metacal_Sheldon2017}. A synthetic pure-noise image is generated
with a different realisation but the same statistical properties as the
observation, rotated by $90^\circ$, and added to the data; in the
Gaussian-smoothed pixel response the two noise contributions then enter
with opposite signs and cancel in expectation. We denote the Fourier
transform of the rotated noise image by $n(\vk)$ and the PSF rotated by
$90^\circ$ by $p^{90}(\vk)$; the latter is used to deconvolve the
injected noise, while the data are deconvolved by the original PSF
$p(\vk)$ and both are reconvolved with the Gaussian kernel
$h(\vk)$. The Fourier transform of the resulting renoised, smoothed
image is
\begin{equation}
\label{eq:smoothed_image}
f_{h}(\vk) =  h(\vk) \left(
    \frac{f_{p}(\vk)}{p(\vk)} + \frac{n(\vk)}{p^{90}(\vk)}
\right) \,.
\end{equation}
A limitation of this construction is that it doubles the effective
image noise prior to detection and measurement; ongoing work on
deep-field renoising \citep{deepfield_metacal2023} aims to mitigate
this cost in future \anacal{} releases.

\subsubsection{Analytical Pixel Shear Response}

\citet{Anacal_Li2023} showed that the shear response of each smoothed pixel
value can be written in closed form. The key result is:
\begin{equation}
\label{eq:pixel_response}
\begin{split}
    \frac{\partial f_{h}(\vx)}{\partial \gamma_1}
    &= G_1(\vx) + x_1 J_1(\vx) - x_2 J_2(\vx)\,, \\
    \frac{\partial f_{h}(\vx)}{\partial \gamma_2}
    &= G_2(\vx) + x_2 J_1(\vx) + x_1 J_2(\vx)\,,
\end{split}
\end{equation}
where the auxiliary filtered images $G_1$, $G_2$, $J_1$, and $J_2$ are linear
functions of the PSF-deconvolved image, defined through Fourier-space operations
\citep{Anacal_Li2023, Anacal_Li2025}:
\begin{equation}
\label{eq:aux_images}
\begin{split}
    G_1(\vx) &= \mathcal{F}^{-1}\!\left[\left(
        \frac{f_p(\vk)}{p(\vk)} - \frac{n(\vk)}{p^{90}(\vk)}
    \right) h(\vk)\,(k_1^2 - k_2^2) \right], \\[4pt]
    G_2(\vx) &= \mathcal{F}^{-1}\!\left[\left(
        \frac{f_p(\vk)}{p(\vk)} - \frac{n(\vk)}{p^{90}(\vk)}
    \right) h(\vk)\,(2k_1 k_2) \right], \\[4pt]
    J_1(\vx) &= \rmi\,\mathcal{F}^{-1}\!\left[\left(
        \frac{f_p(\vk)}{p(\vk)} - \frac{n(\vk)}{p^{90}(\vk)}
    \right) h(\vk)\,k_1 \right], \\[4pt]
    J_2(\vx) &= \rmi\,\mathcal{F}^{-1}\!\left[\left(
        \frac{f_p(\vk)}{p(\vk)} - \frac{n(\vk)}{p^{90}(\vk)}
    \right) h(\vk)\,k_2 \right],
\end{split}
\end{equation}
where $\mathcal{F}^{-1}$ denotes the inverse Fourier transform and
$(k_1,k_2)$ are the Cartesian components of the wave vector $\vk$. The
$G_{1,2}$ images encode the response to shear-induced stretching of the galaxy
profile along the $\gamma_1$ and $\gamma_2$ directions, while $J_{1,2}$ encode
the gradient of the profile (multiplied by $\rmi$), which enters through the
accompanying shift of the distortion reference point.
\citet{Anacal_Li2025b} subsequently extended this analytical pixel-level
shear response to model-fitting shear estimators by introducing
\emph{quintuple numbers}---a generalisation of the dual numbers used
in automatic differentiation, with four nilpotent infinitesimals that
carry the two shear-component derivatives together with the two
reference-point shift derivatives associated with $J_{1,2}$---which
propagate the pixel-level shear response automatically through the
otherwise intractable chain rule of an iterative galaxy model fit.

\subsubsection{Shear Estimator via Chain Rule}
\label{subsubsec:chain_rule}

Given the pixel-level shear responses from equation~\eqref{eq:pixel_response},
the shear response of any observable $O$ measured from the smoothed image can
be computed via the chain rule:
\begin{equation}
    \frac{\partial O}{\partial \gamma_\alpha} = \sum_\vx
    \frac{\partial O}{\partial f_h(\vx)}
    \frac{\partial f_h(\vx)}{\partial \gamma_\alpha}\,, \quad
    \alpha \in \{1, 2\}\,.
\end{equation}
In \citet{Anacal_Li2023}, this framework was used to derive the shear response
of \FPFS{} shapes \citep{FPFS_Li2018, FPFS_Li2022}---a fixed-kernel shape
estimator based on polar shapelets
\citep{shapeletsI_Refregier2003, polar_shapelets_Massey2005}. The \FPFS{}
ellipticity $\epsilon_{1,2}$ and the detection weight $w_d$ are both treated
as differentiable functions of the smoothed image, enabling their shear
responses to be computed analytically.

We define the mean shear response over the detected galaxy sample as
\begin{equation}
\label{eq:response}
    \mathcal{R}_\alpha \equiv
    \left\langle \frac{\partial e_\alpha}{\partial \gamma_\alpha} \right\rangle\,,
\end{equation}
where $e_\alpha = \epsilon_\alpha w_d$ is the weighted ellipticity and
$\langle \cdot \rangle$ denotes averaging over detected galaxies. The
detection weight $w_d$ is a smooth, differentiable proxy for the hard
detection cut: it is constructed from smooth-step functions of the linear
observables, so that the cut transitions continuously from $0$ (rejected)
to $1$ (kept) over a finite width \citep{Anacal_Li2023}. Because $w_d$
is therefore a differentiable function of the smoothed image,
$\mathcal{R}_\alpha$ automatically captures the shear response of both
shape measurement and galaxy detection at second order in shear
\citep{Anacal_Li2023}. The base shear estimator is
\begin{equation}
\label{eq:shear_estimator}
    \hat{\gamma}_\alpha = \frac{\langle e_\alpha \rangle}{\mathcal{R}_\alpha}\,.
\end{equation}
In Section~\ref{subsec:pz_selection}, we extend this estimator to account
for the additional selection imposed by \pz{}-based tomographic binning.

\subsubsection{Shapelet-based Ellipticity and Flux}
\label{subsubsec:fpfs}

We use polar-shapelet observables \citep{shapeletsI_Refregier2003,
polar_shapelets_Massey2005} as the building blocks of
both the \FPFS{} shape and the \anacal{} flux entering the \pz{} estimator.
Each shapelet mode $M_{nm}$ is the projection of the smoothed,
PSF-deconvolved image $f_h(\vx)$ onto the polar shapelet basis function
$\phi_{nm}(\vx \,|\, \sigma_h)$ of radial number $n$ and spin $m$,
\begin{equation}
\label{eq:shapelet_mode}
    M_{nm} = \int f_h(\vx)\, \phi_{nm}^{*}(\vx \,|\, \sigma_h)\, \rmd^2 \vx\,,
\end{equation}
where $\sigma_h$ is the Gaussian smoothing scale of
equation~\eqref{eq:smoothed_image}. Following \citet{Anacal_Li2024} we
adopt $\sigma_h = 0\farcs56$, which approximately maximises the
effective galaxy number density on LSST-like data.

The spin-2 \FPFS{} ellipticity \citep{FPFS_Li2018, FPFS_Li2022} is
constructed from the $n=2$ shapelet modes,
\begin{equation}
\label{eq:fpfs_ellipticity}
    \epsilon_1 + \rmi\, \epsilon_2 \equiv \frac{M_{22}}{M_{00}+C}\,,
\end{equation}
where $C$ is a sample-dependent weighting parameter introduced by
\citet{FPFS_Li2018} that controls the relative weight of bright versus faint
galaxies. We adopt $C = 4\sigma_0$, following \citet{Anacal_Li2025},
where $\sigma_0$ is the standard deviation of the measurement noise on
$M_{00}$. The $\epsilon_{1,2}$ in equation~\eqref{eq:fpfs_ellipticity} are
the spin-2 components entering the shape factor $e_{\alpha}=\epsilon_{\alpha}
w_d$ in Section~\ref{subsubsec:chain_rule}.

For the \pz{} estimators, we use the zeroth-order shapelet mode $M_{00}$ as
the \anacal{} flux. Because $\phi_{00}$ is a Gaussian of width $\sigma_h$,
$M_{00}$ is a Gaussian-aperture flux measured on the same PSF-rounded,
renoised image that defines the shear response in
equations~\eqref{eq:pixel_response}--\eqref{eq:aux_images}.
In this sense the \anacal{} flux is essentially the same
class of measurement as the Gaussian-Aperture-and-PSF (GAaP) photometry
\citep{kuijken2008} used in the LSST Science Pipelines
\citep{kannawadi2022}: both are Gaussian,
PSF-homogenised flux measurements evaluated on images reconvolved to a
common, PSF-rounded resolution, so that the aperture scale is matched
across bands and the resulting colours are consistent. The distinction is
technical: \anacal{} realises this measurement through the zeroth-order
polar-shapelet mode $M_{00}$, which corresponds to a Gaussian-aperture flux
at the specific image resolution set by the reconvolution kernel. The per-band
\anacal{} fluxes used as input to the \pz{} algorithms in
Section~\ref{sec:dp1} are therefore the $\{M_{00}^{(b)}\}_{b\in\{u,g,r,i,z,y\}}$
modes measured on each filter. Sharing the same smoothed image between the
flux measurement and the shear response enables the \pz{}-based
tomographic selection to be propagated through the analytical chain rule of
Section~\ref{subsec:pz_selection}, without an additional simulation-based
calibration step.

In its current form, the \FPFS{} estimator described above does not
include an explicit deblending step: each detected peak is treated as an
isolated source, so the $M_{00}$ fluxes used as \pz{} inputs and the
$M_{22}$ shapes used for shear inherit any flux contamination from
overlapping neighbours. We plan to incorporate deblending into the
\anacal{} framework using the deep-learning-based estimator of
\citet{Lin2026_D4CNN_AnaCal}, in which a physics-informed convolutional
network ($D_4$CNN) is combined with the \anacal{} response, and to
validate the resulting tomographic shear calibration on blended image
simulations.

\subsection{Photo-$z$ Selection Bias Correction}
\label{subsec:pz_selection}

Tomographic weak lensing assigns source galaxies to redshift bins using
photo-$z$ point estimates derived from galaxy fluxes, or using self-organising
map (SOM) cells trained on the same fluxes. Galaxy $i$ is placed in bin
$t$ if its point-estimate redshift $\hat{z}_i \in [z_\mathrm{low}^{(t)},\,
z_\mathrm{high}^{(t)})$. Since galaxy flux and colour measurements are
correlated with the underlying shear distortion
(Section~\ref{subsec:anacal_review}), the tomographic bin membership
is implicitly shear-dependent. As a result, the
mean shear response $\mathcal{R}_\alpha$
(equation~\ref{eq:response}) of the photo-$z$-selected subsample in each bin
generally differs from that of the full detected catalogue; calibrating with
the full-catalogue response therefore introduces a \emph{tomographic selection
bias}.

To correct for this effect, we propagate the analytical \anacal{} flux
responses $\partial F_b / \partial \gamma_\alpha$ through the \pz{} code
following the metacalibration convention \citep{metacal_Sheldon2017}.
We construct shear-perturbed versions of the multiband fluxes using a
finite-difference step $\Delta\gamma = 0.01$:
\begin{equation}
\label{eq:flux_perturb}
    F_{b}^{\pm} = F_{b} \pm \Delta\gamma\,
        \frac{\partial F_{b}}{\partial \gamma_\alpha}\,,
\end{equation}
where $b \in \{u,g,r,i,z,y\}$ indexes the photometric band. We pass
$\{F_b^{\pm}\}$ through the \pz{} estimator to obtain perturbed redshift
estimates $z_\mathrm{mode}^{\pm}$, assign galaxies to tomographic bin $t$ via
hard cuts on $z_\mathrm{mode}^{\pm}$, and compute the mean perturbed weighted
ellipticities $\langle e_\alpha^{+,(t)}\rangle$ and
$\langle e_\alpha^{-,(t)}\rangle$. The selection shear response for bin $t$
is estimated by finite difference:
\begin{equation}
\label{eq:sel_response}
    \mathcal{R}_{\alpha,\mathrm{sel}}^{(t)} =
        \frac{\langle e_{\alpha}^{+,(t)} \rangle -
              \langle e_{\alpha}^{-,(t)} \rangle}
             {2\,\Delta\gamma}\,.
\end{equation}

The tomographic shear estimator with the full selection bias correction is
\begin{equation}
\label{eq:tomo_shear_estimator}
    \hat{\gamma}_\alpha^{(t)} =
        \frac{\langle e_\alpha^{(t)} \rangle}
             {\mathcal{R}_\alpha + \mathcal{R}_{\alpha,\mathrm{sel}}^{(t)}}\,,
\end{equation}
where $\langle e_\alpha^{(t)} \rangle$ is the mean weighted ellipticity of
galaxies in bin $t$ under the original photo-$z$ assignments.  The
denominator combines the \anacal{} shape response $\mathcal{R}_\alpha$
(equation~\ref{eq:response}) with the selection response
$\mathcal{R}_{\alpha,\mathrm{sel}}^{(t)}$ (equation~\ref{eq:sel_response}),
each evaluated separately for every tomographic bin $t$ and shear component
$\alpha$.

\section{PHOTO-$z$ ESTIMATION IN LSST DP1}
\label{sec:dp1}

\begin{figure*}[t]
    \centering
    \includegraphics[width=.95\textwidth]{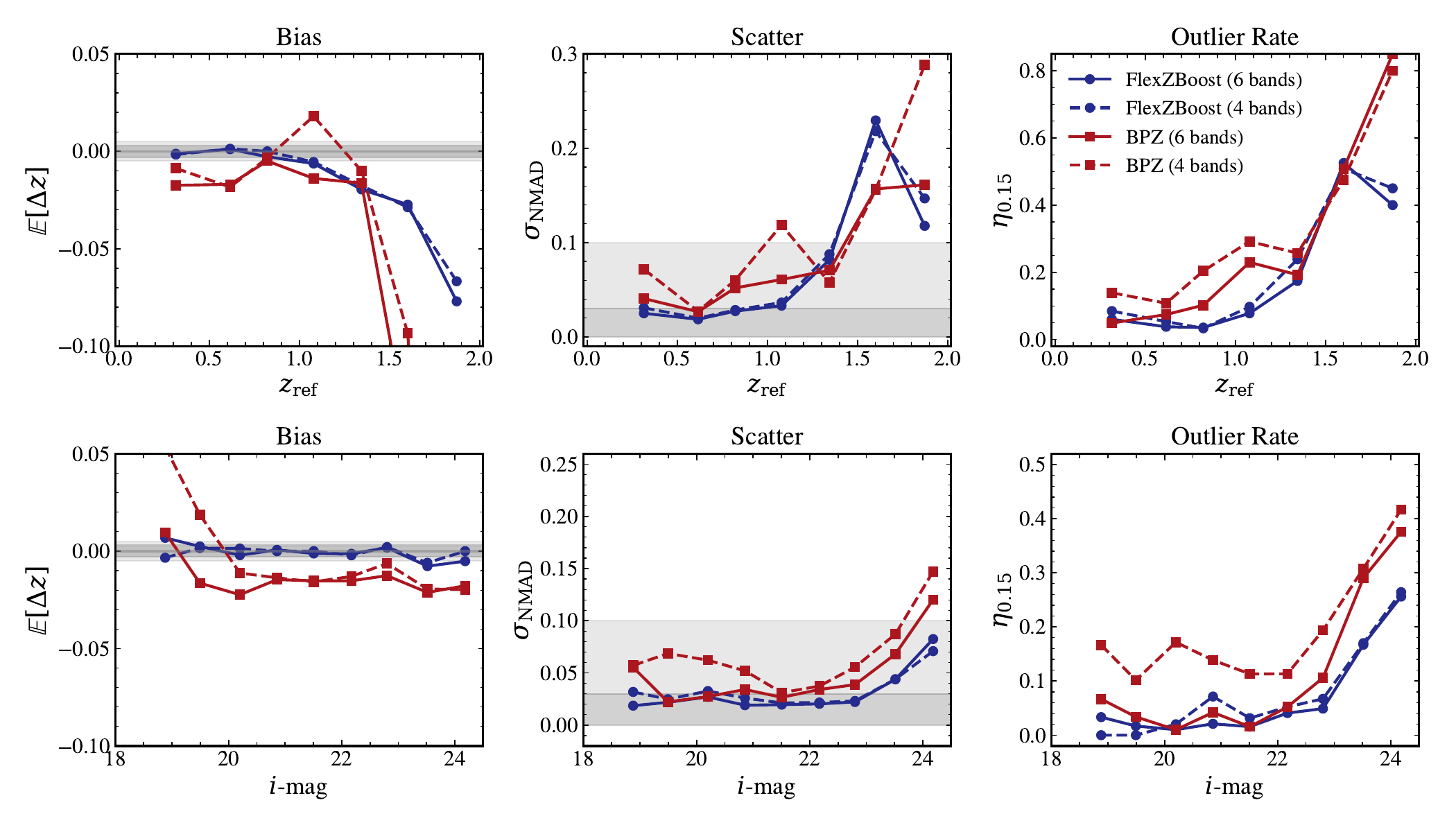}
    \caption{
        Photo-$z$ performance metrics for \FlexZBoost{} (blue circles)
        and \BPZ{} (red squares) on the DP1 \texttt{ECDFS} test set as
        a function of reference spec-$z$ (top row) and $i$-band
        magnitude (bottom row). Each panel shows one of three metrics:
        the photo-$z$ bias $\mathbb{E}[\Delta z]$, scatter $\sigmaNMAD$,
        and outlier rate $\eta_{0.15}$. Solid lines show the six-band
        ($ugrizy$) configuration; dashed lines show the four-band
        ($griz$) configuration. The grey shaded regions show the LSST
        Dark Energy Science Collaboration (DESC) Science Requirements
        Document \citep[SRD;][]{LSSTRequirement2018} Y1 (lighter) and Y10
        (darker) requirements on the mean and scatter of the tomographic
        $n(z)$. These are not requirements on the point estimates plotted
        here; we show them only as a reference scale (see text). The \pz{}
        performance starts to degrade when $z > 1.2$ and
        $i$-mag$ > 23$. We do not show statistics beyond
        $z_\mathrm{ref}=2$ because of the limited number of reference
        galaxies. The performance shown in this figure is evaluated on the
        $30\%$ test sample.
    }
    \label{fig:metrics_combined}
\end{figure*}

As a first sanity check on real data, we verify that \pz{} estimates
derived from the \anacal{} fluxes are of comparable quality to those
obtained from the standard LSST photometry pipeline, before turning to
the sim-based validation of the analytical selection-bias correction in
Section~\ref{sec:validation}. This establishes that the same flux
measurements that feed the \anacal{} shear estimator can also be used as
input to the \pz{} estimator without degrading photometric-redshift
performance.
Section~\ref{subsec:dp1_data} describes the DP1 photometric and reference
samples; Section~\ref{subsec:pz_performance_dp1} presents the \pz{}
performance metrics on the DP1 \texttt{ECDFS} test set.

\subsection{Rubin Observatory Data Preview 1}
\label{subsec:dp1_data}

The Rubin Observatory Data Preview 1 \citep[DP1;][]{dp1_dataset,
dp1_paper} is the first
real data release from the NSF/DOE Vera C.\ Rubin Observatory, and is based
on commissioning observations taken with the LSST Commissioning Camera
(LSSTComCam, \citealt{SITCOMTN-149}) mounted on the 8.4-meter Simonyi Survey
Telescope. LSSTComCam consists of a single raft of nine 4K$\times$4K CCDs
covering a field of view of $\sim 40^{\prime}\times 40^{\prime}$, much
smaller than the 21-raft, 189-CCD, 3.2-gigapixel full LSSTCam, which
spans $9.6\,\mathrm{deg}^2$. The on-sky campaign ran from 2024-10-24 to
2024-12-11 and obtained imaging across seven fields distributed in the
southern sky, covering approximately $15\,\mathrm{deg}^2$ in up to six
optical bands ($ugrizy$). The raw data
were processed with the LSST Science Pipelines
\citep{PSTN-019}, which perform instrument
signature removal, astrometric and photometric calibration, PSF
modelling, and image coaddition.

Of the seven DP1 fields, three---the Extended Chandra Deep Field South
(\texttt{ECDFS}), the Euclid Deep Field South (\texttt{EDFS}), and the
Low Galactic Latitude Field---have full six-band ($ugrizy$) coverage;
the Low Ecliptic Latitude Field (\texttt{Rubin\_SV\_38\_7}) was imaged
in only the four $griz$ bands, and the remaining fields cover other
subsets of the LSST filters \citep{RTN-095}. Among the three six-band
fields, \texttt{ECDFS} is the only one with an extensive reference
redshift catalogue, built from spectroscopic, grism, and multiband
\pz{} surveys overlapping the DP1 footprint (described in detail
below). We therefore adopt \texttt{ECDFS} for both training and
testing of the \pz{} estimators in this work. Characterising
the \pz{} performance on \texttt{ECDFS} with both the full six-band
photometry and a reduced four-band ($griz$) configuration brackets the
range of \pz{} quality expected across the DP1 footprint and informs
what to expect when the same algorithm is applied to fields without
$u$- and $y$-band coverage, such as the Low Ecliptic Latitude Field.
Galaxy detection and shear measurement are performed on the $i$-band
coadds, while the \anacal{} fluxes used for the \pz{} estimation are
measured in all available bands.
We apply the standard quality cuts summarised in Table~\ref{tab:cuts},
comprising an $i$-band \anacal{} flux signal-to-noise cut, per-band
magnitude limits on the Gaussian-aperture \anacal{} fluxes, and a cut on
the trace of the $i$-band \FPFS{} second-order moments.

\begin{table}
    \centering
    \caption{Quality cuts applied to the DP1 source catalogue. All magnitude
        limits refer to the Gaussian-aperture \anacal{} fluxes with magnitude
        zeropoint $31.4$.}
    \label{tab:cuts}
    \begin{tabular}{lc}
        \hline
        Quantity & Cut \\
        \hline
        $i$-band \anacal{} flux SNR & $>5$ \\
        $u$-band magnitude & $<27.5$ \\
        $g$-band magnitude & $<26.5$ \\
        $r$-band magnitude & $<26.0$ \\
        $i$-band magnitude & $<25.0$ \\
        $z$-band magnitude & $<25.0$ \\
        $y$-band magnitude & $<25.0$ \\
        $i$-band moment trace $(M^{(i)}_{00} + M^{(i)}_{20})/M^{(i)}_{00}$ & $>0.10$ \\
        \hline
    \end{tabular}
\end{table}

The input fluxes used to train and run
the \pz{} estimators are the \anacal{} fluxes, which are measured together with
their analytical pixel shear responses (Section~\ref{subsec:anacal_review}).
The same flux measurements therefore feed into both the shear estimator and
the \pz{} estimator, ensuring that the selection bias correction
described in Section~\ref{subsec:pz_selection} is fully self-consistent.

The reference catalogue used to train and calibrate the \pz{} algorithms is
assembled from spectroscopic, grism, and multiband \pz{} surveys spanning more
than two dozen programs \citep{dp1-pz}. After cross-matching to the DP1 object
catalogue and requiring all-band detections with $i$-band CModel signal-to-noise
ratio $> 20$ and redshift confidence $> 0.9$, 8314 galaxies remain, which
we split 70/30 into training (5819 galaxies) and test (2495 galaxies)
sets, which are used to train \FlexZBoost{} and to calibrate \BPZ{}.
The average confidence of the training sample is $0.97$, indicating
approximately $3\%$ redshift outliers in the training set.

%\rachel{Are you using some default parameter set for these algorithms?  If so, I think you should say that.  Otherwise, need to document what was used.}

Both algorithms are run through their \RAIL{} implementations with the
configurations summarised in Table~\ref{tab:pz_config}. For \BPZ{} we use a
library of 136 SED templates built on the LePhare base SEDs of
\citet{Ilbert:09} with Calzetti-like dust extinction added to the later-type
templates, partitioned into 7 elliptical, 57 spiral, and 72
starburst/irregular templates for the type-and-magnitude prior of
\citet{Benitez:2000}, which is enabled (\texttt{no\_prior=false}). For
\FlexZBoost{} \citep{Izbicki:2017} we use the \RAIL{} default 35 basis
functions, train on the full training set (\texttt{trainfrac}$=1$), and adopt
a single point on the bump-removal and PDF-sharpening grids
(\texttt{bumpthresh}$=0.05$, \texttt{sharpness}$=0.8$); all other
hyper-parameters are left at their \RAIL{} defaults.

\begin{table}
    \centering
    \caption{\RAIL{} configuration of the two \pz{} algorithms used in
        this work. Parameter names match the \RAIL{} \texttt{yaml} keys.}
    \label{tab:pz_config}
    \begin{tabular}{lll}
        \hline
        Algorithm & Parameter & Value \\
        \hline
        \BPZ{}        & Number of SED templates & 136 \\
                      & \texttt{nt\_array}     & $[7,\,57,\,72]$ \\
                      & \texttt{no\_prior}     & \texttt{false} \\
        \hline
        \FlexZBoost{} & \texttt{max\_basis}    & 35 \\
                      & \texttt{trainfrac}     & 1.0 \\
                      & \texttt{bumpmin}, \texttt{bumpmax}, \texttt{nbump} & $0.05,\,0.07,\,1$ \\
                      & \texttt{sharpmin}, \texttt{sharpmax}, \texttt{nsharp} & $0.8,\,1.1,\,1$ \\
        \hline
    \end{tabular}
\end{table}

\subsection{Photo-$z$ Performance}
\label{subsec:pz_performance_dp1}

We characterise the \pz{} point-estimate quality of \FlexZBoost{} and
\BPZ{} on the DP1 \texttt{ECDFS} test set by computing the mean bias
$\mathbb{E}[\Delta z]$, scatter $\sigmaNMAD$, and catastrophic outlier
rate $\eta_{0.15}$ (Section~\ref{subsec:rail}) in bins of reference
spec-$z$ and $i$-band magnitude. For each algorithm we compare two flux
configurations: the default six-band ($ugrizy$) \anacal{} fluxes and a
reduced four-band ($griz$) configuration that omits the $u$ and $y$ bands. The four-band setup
uses the $griz$ \anacal{} fluxes measured on \texttt{ECDFS} to mimic
the photometric coverage of the Low Ecliptic Latitude Field
(Section~\ref{subsec:dp1_data}), so that the resulting metrics
indicate the \pz{} quality expected when the same algorithm is applied
to that field.

Figure~\ref{fig:metrics_combined} shows the three metrics as a function
of reference spec-$z$ (top row) and $i$-band magnitude (bottom row).
Across the full redshift and magnitude ranges, \FlexZBoost{} delivers
smaller bias, scatter, and outlier rate than \BPZ{}, consistent with
the relative ordering of template-fitting and machine-learning
algorithms reported in \citet{dp1-pz}. The performance is best at
intermediate redshift ($0.3 < z < 1.2$) and bright magnitudes
($i < 23$) for both algorithms, and degrades at $z > 1.2$ and
$i$-mag$ > 23$. For reference, the grey bands in
Figure~\ref{fig:metrics_combined} show the LSST Dark Energy Science
Collaboration (DESC) Science Requirements Document (SRD) Y1 (lighter) and
Y10 (darker) requirements on the inferred tomographic $n(z)$
\citep{LSSTRequirement2018}: the systematic uncertainty in the mean redshift
of each bin must not exceed $0.005\,(1+z)$ at Y1 and $0.003\,(1+z)$ at Y10,
and the systematic uncertainty in the photometric redshift scatter $\sigma_z$
must not exceed $0.1\,(1+z)$ at Y1 and $0.03\,(1+z)$ at Y10. These are
requirements on the ensemble $n(z)$ per tomographic bin, not on the
per-galaxy point estimates plotted here; we show them as a reference scale
only. With this caveat, the $\sigmaNMAD$ of the \FlexZBoost{} six-band
configuration lies within the Y1 reference band for $z < 1.2$, while
$\mathbb{E}[\Delta z]$ exceeds it in several bins for both algorithms,
underscoring the importance of ensemble $n(z)$ calibration via techniques
such as clustering redshifts \citep{Newman08}.

The two algorithms also respond differently to dropping the $u$ and
$y$ bands. For \BPZ{}, switching from the six-band to the four-band
configuration noticeably degrades all three metrics, with the
largest impact on the outlier rate $\eta_{0.15}$, where the
additional $u$ and $y$ coverage helps break colour--redshift
degeneracies in the template fits. For \FlexZBoost{}, by contrast,
the six- and four-band curves nearly overlap across both $z_{\rm ref}$
and $i$-mag, indicating that the machine-learning estimator is much
less sensitive to the loss of $u$ and $y$ on \texttt{ECDFS}; this
suggests that \FlexZBoost{} should retain comparable point-estimate
performance on the four-band DP1 fields, whereas \BPZ{} will be
appreciably degraded there.

Compared with the corresponding analysis of \citet[Figure~6]{dp1-pz},
which evaluates the same algorithms with the standard LSST GAaP
fluxes, our \pz{} estimates derived from the \anacal{} fluxes show a
smaller bias at high redshift for faint galaxies (left panel of
Figure~\ref{fig:metrics_combined}), while the scatter $\sigmaNMAD$
and outlier rate $\eta_{0.15}$ are comparable to those reported in
\citet{dp1-pz}. This indicates that using \anacal{} fluxes as the
\pz{} input does not degrade \pz{} performance relative to the GAaP
baseline and, in the regime that dominates the high-redshift
tomographic bins, slightly improves the mean redshift bias.
We caution, however, that this is not a fully controlled,
object-by-object comparison: the detection catalogue and the sample
selection underlying our \anacal{}-flux analysis are not identical to those
of \citet{dp1-pz}. The modest improvement in the high-redshift mean bias
should therefore be read as evidence that \anacal{} fluxes perform at least
as well as the GAaP baseline, rather than as a robust demonstration that
they are more accurate. Having
shown on real data that \anacal{} fluxes yield \pz{} estimates of
comparable or better quality than the GAaP baseline, we now turn to
the central test of this paper: whether the \pz{}-based tomographic
shear estimate is accurately calibrated by the \anacal{} analytical
selection-bias correction on LSST-like image simulations.

\section{VALIDATION ON LSST-LIKE IMAGE SIMULATIONS}
\label{sec:validation}

We now validate the \anacal{} \pz{} selection bias correction on LSST-like
image simulations with blending at the expected LSST Y10 depth. Section~\ref{subsec:sim_setup}
describes the simulation setup and the motivation for the Y10 depth choice;
Section~\ref{subsec:bias_measurement} reports the measured multiplicative
shear bias before and after correction for both \pz{} algorithms.

\subsection{Simulation Setup}
\label{subsec:sim_setup}

To validate the \anacal{} selection bias correction described in
Section~\ref{subsec:pz_selection}, we use the image simulation framework from
\citet{metaDet_LSST2023}, which is built on the \galsim{} package
\citep{GalSim} and generates realistic synthetic galaxy images with LSST-like
observing conditions. We simulate coadded images in the $i$~band under the expected
LSST observing conditions, with a Moffat PSF of FWHM $0\farcs80$, a pixel
scale of $0\farcs20$, and a noise level corresponding to the expected LSST
Y10 depth, following \citet{metaDet_LSST2023} and the LSST DESC Science
Requirements Document \citep{LSSTRequirement2018}.

The simulation includes blended galaxy systems and applies a uniform shear
distortion of $\gamma_1 = \pm 0.02$ across each subfield. We generate 60{,}000
subfield pairs, each covering $0.036\,\mathrm{deg}^2$, for a total
simulated area of $\sim 2{,}160\,\mathrm{deg}^2$, and apply the shape
noise cancellation technique of \citet{preciseSim_Pujol2019} to suppress
statistical noise in shear bias estimation. Source galaxies are drawn from the CatSim-based models
\citep{connollySimulatingLSSTSystem2010,
connollyEndtoendSimulationFramework2014} used in LSST DESC Data Challenge 1
\citep{sanchezLSSTDESCData2020} and calibrated to Hyper Suprime-Cam data
\citep{aiharaSecondDataRelease2019}. The catalogue has a raw source density of
$\sim 240~\mathrm{arcmin}^{-2}$, and an effective source number density of
$n_{\rm eff} \sim 17 \; \mathrm{arcmin}^{-2}$
\citep{parkDeepfieldAnalyticalCalibration2026, Anacal_Li2025} after
detection and selection; this number density corresponds to
detection and shear measurement on a single band, while the four-band
measurement gives an effective number density of $n_{\rm eff} \sim 26
\; \mathrm{arcmin}^{-2}$. The galaxy models include bulge, disk, and AGN
components, and reach an effective $i$-band AB magnitude limit of $\sim 27$.

\begin{figure}
    \centering
    \includegraphics[width=.45\textwidth]{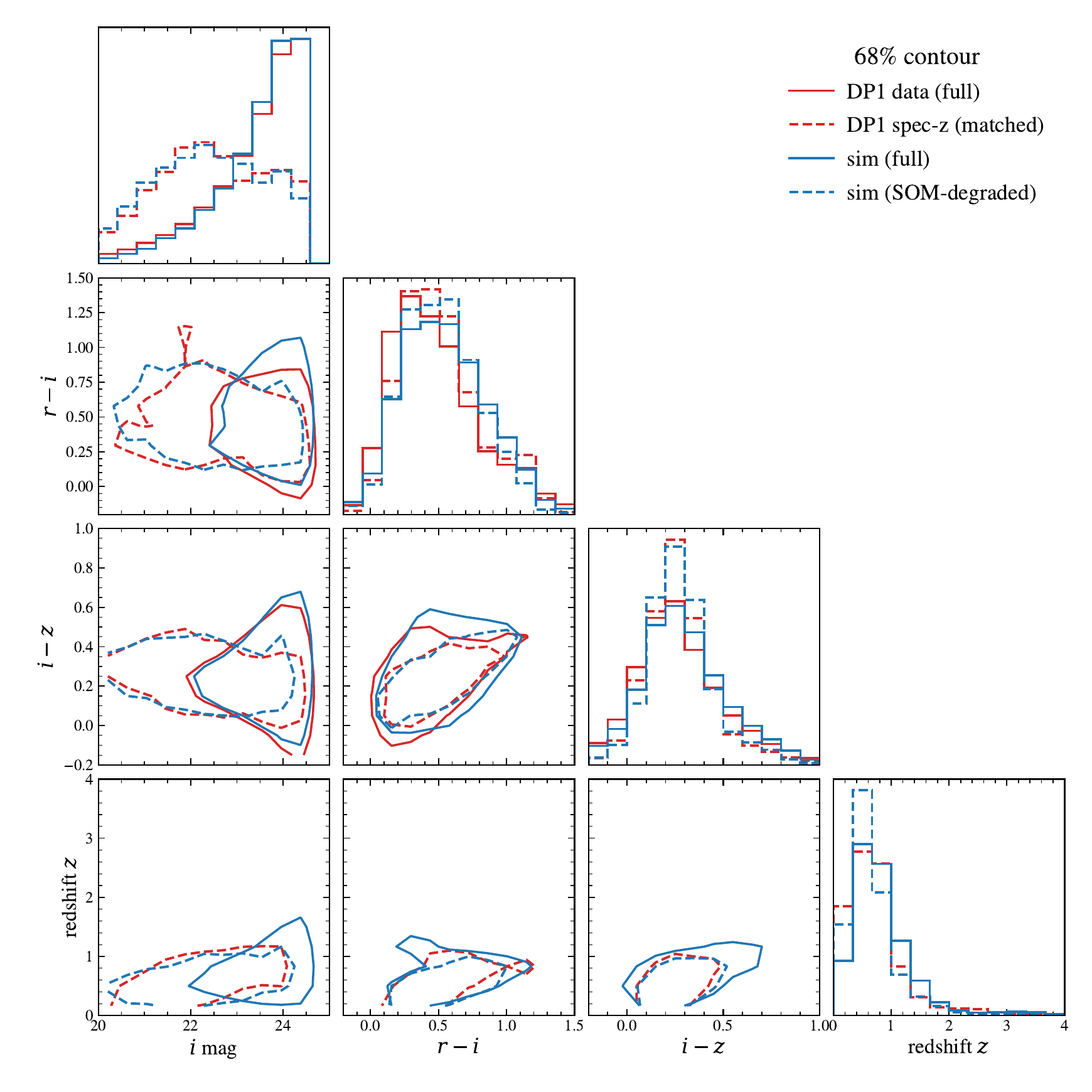}
    \caption{
        Distribution of four samples in the $(i, r-i, i-z, \mathrm{redshift})$
        space, restricted to $i < 24.5$: the DP1 photometric sample in the
        \texttt{ECDFS} field (red solid) and its spec-$z$ matched subset (red
        dashed), together with the full LSST-like simulation photometric sample
        (blue solid) and the SOM-degraded spec-$z$ training sample drawn from
        it (blue dashed). Diagonal panels show the marginal density of each
        variable; off-diagonal panels show contours enclosing 68\% of the 2-D
        density. The simulation degraded sample is constructed with the SOM
        degrader in \RAIL{} using the \texttt{ECDFS} spectroscopic sample
        overlapping DP1 as the reference (see Section~\ref{subsec:dp1_data} for
        further details on the DP1 spec-$z$ sample).
    }
    \label{fig:color_mag_contours_dp1}
\end{figure}

In real observations, spectroscopic redshift surveys are far shallower and
sparser than the photometric samples to which the \pz{} pipeline is ultimately
applied. The spec-$z$ training set is therefore a non-representative subset of
the photometric pool, preferentially populated by bright galaxies whose
redshifts are comparatively easy to measure. Training and validating \pz{}
estimators on such a sample risks overestimating their accuracy on the full
photometric population that the pipeline must process in production.
%\rachel{Optimistic in what way?}
To emulate this regime in our simulation, we
construct
a photometric training sample whose colour--magnitude occupancy
matches that of a realistic spec-$z$ calibration set by applying
the SOM-based ``spec-$z$ degrader'' from \RAIL{}. The degrader
trains a self-organising map (SOM) on the colours and spec-$z$
of a spec-$z$ reference---here a $10\times10$ grid trained
for $10$ epochs on the $(u, g, r, i, z, y)$ colours, and spec-$z$ of the DP1
\texttt{ECDFS} spectroscopic sample (Section~\ref{subsec:dp1_data})---assigns
every source in the wider photometric pool to its best-matching SOM cell, and
sub-samples the pool per cell so that the resulting cell occupancies reproduce
those of the spec-$z$ reference. Because the degrader is driven by the real DP1
\texttt{ECDFS} spec-$z$ sample, the resulting training set inherits the
spec-$z$ selection function in colour--magnitude space and allows us to test the
\pz{} pipeline on simulated data that mimic the selection biases of a real
spec-$z$ calibration set. The resulting colour--magnitude and
redshift distribution of the
degraded spec-$z$ training sample, together with that of the underlying
photometric pool, is shown in Figure~\ref{fig:color_mag_contours_dp1}.
In both the real DP1 data and the simulation, the spec-$z$ contours
(dashed) are systematically offset towards brighter $i$-band magnitudes
relative to the underlying photometric pools (solid), reflecting the
preferential coverage of bright sources by spectroscopic surveys. Two
quantitative comparisons support the fidelity of the simulated training
sample. First, the dashed red (DP1 spec-$z$ matched) and dashed blue
(simulation SOM-degraded) contours closely overlap, showing that the
simulated training sample reproduces the colour--magnitude and
redshift distributions of
the real spec-$z$ sample. Second, the two solid curves also agree, confirming
that the underlying photometric pools are well matched between the
simulation and DP1. Together, these comparisons show that our simulation
faithfully captures both the photometric pool and the non-representative
spec-$z$ subset drawn from it.

%\rachel{I don't understand how this is a
%quantitative confirmation that we reproduce real selection biases when there is
%no real sample shown on the plot.}

Both \pz{} algorithms are run with the same \RAIL{}
configurations as in Section~\ref{sec:dp1} (Table~\ref{tab:pz_config}):
\FlexZBoost{} is retrained on the SOM-degraded simulation training sample
described above, while \BPZ{} uses the same template library of
\citet{Ilbert:09} without modification. Although the simulated galaxy SEDs
are generated from the spectral synthesis models of \citet[hereafter
BC03]{bc03} rather than drawn from real spectra, the template
library---empirical COSMOS templates supplemented by BC03-based starburst
models---spans a similar SED space, so the templates provide a reasonable
description of the simulated population. More importantly, the goal of this
section is to test whether the analytical correction removes the shear
selection bias induced by \pz{}-based binning: the tomographic selection
defined by any \pz{} estimate constitutes a valid selection for this test,
so the fidelity of the templates to the simulated population affects the
\pz{} precision but not the validity of the selection-bias validation.

\subsection{Validation of Selection Bias Correction}
\label{subsec:bias_measurement}

We adopt the expected LSST Y10 depth because the LSST DESC SRD shear
calibration requirement $|m| < 3\times10^{-3}$ \citep{LSSTRequirement2018}
is defined for the full ten-year survey, where statistical uncertainties on
cosmological parameters are smallest and the tolerance for residual
systematics is correspondingly tightest. Validating the selection bias
correction at Y10 depth therefore probes the regime in which the
requirement must be satisfied.

To quantify the impact of \pz{}-based binning on shear calibration, we
measure the multiplicative bias $m$ from the \pz{} selection using the
standard estimator \citep{shearSys_Huterer2006, Heymans2006}:
\begin{equation}
\label{eq:mbias}
    \hat{\gamma}_\alpha^{(t)} = (1 + m^{(t)})\,\gamma_\alpha + c^{(t)}_\alpha\,,
\end{equation}
where $m^{(t)}$ and $c^{(t)}_\alpha$ are the multiplicative and additive shear
biases for tomographic bin $t$. We evaluate these biases separately for each of
the five tomographic bins and for both shear components. The five tomographic
bins are defined by binning the mode redshift estimate $z_\text{mode}$ with
edges $\{0.3, 0.6, 0.9, 1.2, 1.5, 1.8\}$.

\begin{figure}
    \centering
    \includegraphics[width=.45\textwidth]{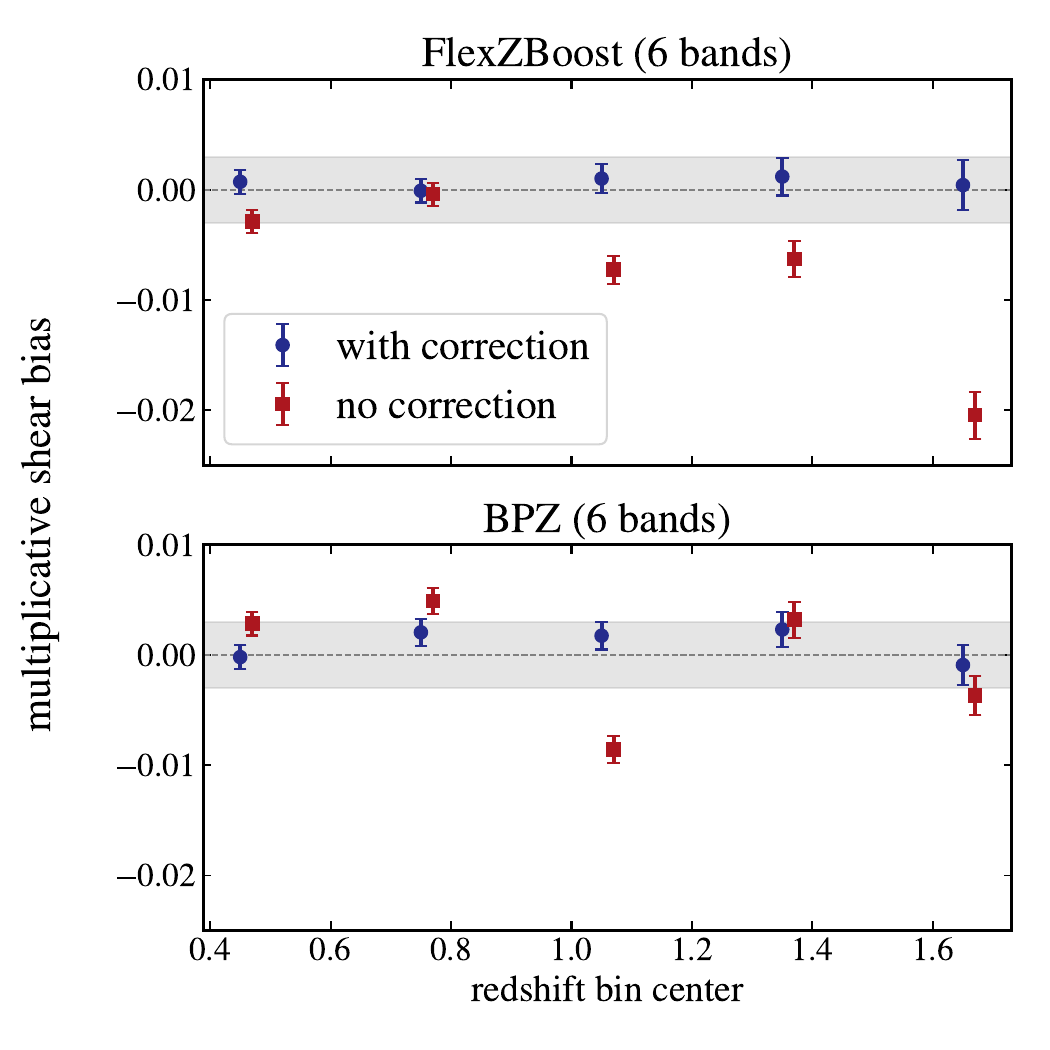}
    \caption{
        Multiplicative shear bias $m$ for each tomographic bin for
        \FlexZBoost{} (upper panel) and \BPZ{} (lower panel), using six-band
        ($ugrizy$) photometry to assign galaxies to bins. Red squares show the
        uncorrected estimator; blue circles show the result after applying the
        \anacal{} selection bias correction
        (equation~\ref{eq:tomo_shear_estimator}). The grey shaded band
        indicates the LSST Y10 requirement $|m| < 3\times10^{-3}$. The
        uncorrected estimator is never applied in practice and is shown only to
        illustrate the size of the tomographic selection bias, which reaches
        $\sim 2\%$ for \FlexZBoost{} and $\sim 1\%$ for \BPZ{}. The corrected
        estimator, whose selection response is derived analytically from the
        data themselves rather than from external image simulations, is within
        the Y10 requirement for all five bins and both algorithms.
    }
    \label{fig:bias_6bands}
\end{figure}

\begin{figure}
    \centering
    \includegraphics[width=.45\textwidth]{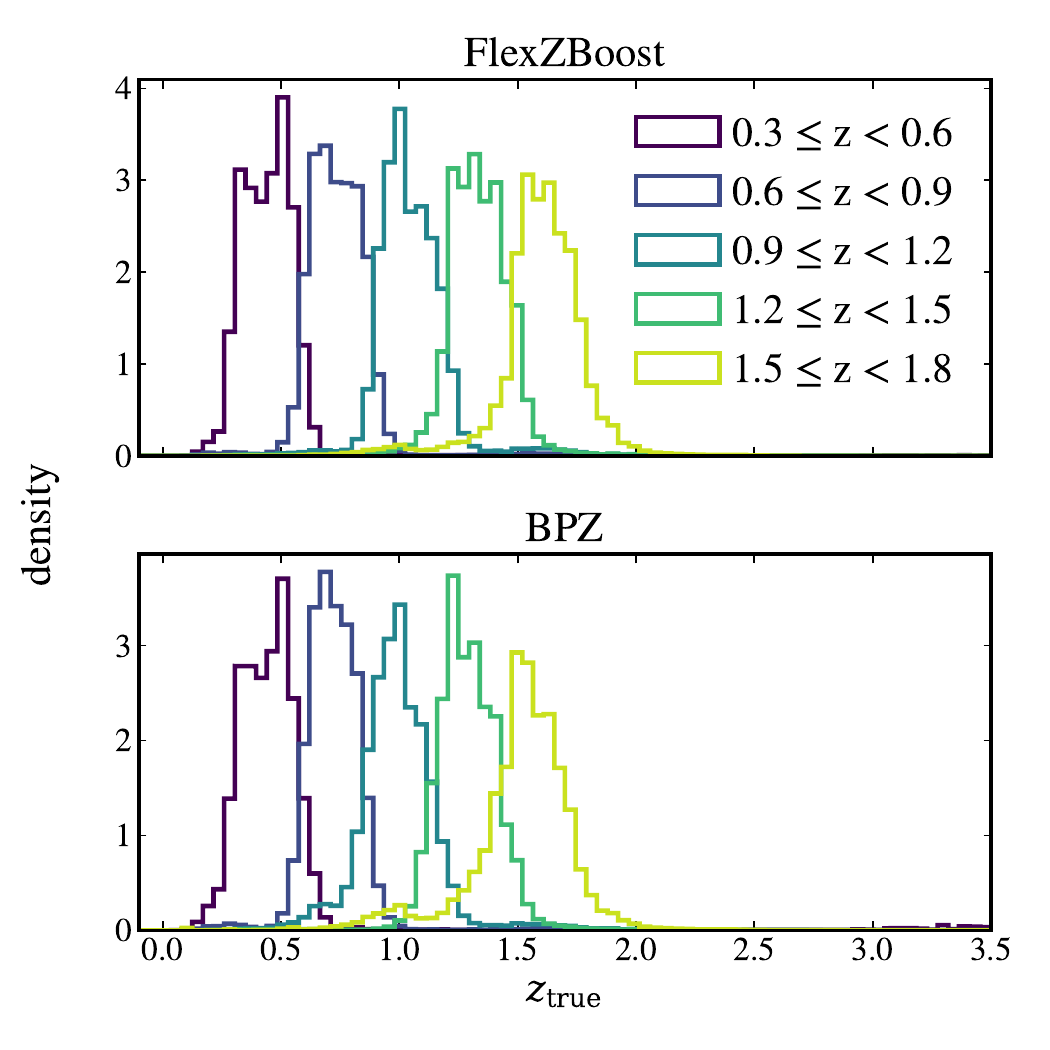}
    \caption{
        Estimated redshift distributions $n(z)$ of the five tomographic
        bins on the LSST-like image simulation at Y10 depth, for
        \FlexZBoost{} (upper panel)
        and \BPZ{} (lower panel). Bin assignment is made by a hard cut on
        the photo-$z$ point estimate derived from the six-band ($ugrizy$)
        \anacal{} fluxes. The two algorithms produce qualitatively
        similar tomographic distributions, with \BPZ{} showing somewhat
        broader tails and a larger fraction of catastrophic outliers in
        the high-redshift bins.
    }
    \label{fig:nz_sim}
\end{figure}

Figure~\ref{fig:bias_6bands} shows the resulting multiplicative bias for
\BPZ{} and \FlexZBoost{} across the five tomographic bins, before and after
the \anacal{} selection bias correction. Without the correction, the bias
exceeds the Y10 requirement in most bins; the largest deviation occurs in
the highest redshift bin, where the \pz{} outlier fraction peaks. After
applying the correction, the bias falls within the requirement for all
five bins and for both algorithms. The corresponding estimated tomographic
redshift distributions are shown in Figure~\ref{fig:nz_sim}. The $n(z)$ in each
tomographic bin is the histogram of the truth redshifts of the galaxies
assigned to that bin, where truth redshifts are obtained by matching the
detected sources to the input catalog within a $0.5$~arcsec radius. The five
bins are cleanly separated in redshift, with each bin dominated by a single
peak at its target redshift and only a small fraction of catastrophic outliers
leaking into neighbouring bins. This indicates that, on our LSST-like
simulation, both \pz{} algorithms are able to place galaxies into the intended
tomographic bins with reasonably low outlier contamination.

Without the \anacal{} analytical selection-bias correction the \pz{}-binned
shear estimate is significantly biased by the selection effect, exceeding
the LSST Y10 requirement $|m| < 3\times10^{-3}$ in most tomographic bins
for both \BPZ{} and \FlexZBoost{}. After applying the
correction---finite differencing of the \pz{} estimates with the
analytical shear responses of the colours---the
residual multiplicative bias is statistically consistent with zero and
lies within the Y10 requirement across all five bins and for both \pz{}
algorithms, demonstrating that the \pz{}-based selection bias can be
removed self-consistently within the \anacal{} framework, without recourse
to external image simulations for calibration.

\section{SUMMARY AND OUTLOOK}
\label{sec:summary}

In this paper, we have extended the \anacal{} analytical shear calibration
framework to correct for selection biases introduced by photometric
redshift-based tomographic binning in weak gravitational lensing analyses,
and validated the \pz{} estimates derived from \anacal{} fluxes on the
Rubin Observatory DP1 \texttt{ECDFS} field. Our main results are as
follows:

\begin{enumerate}
    \item We extend the \anacal{} framework to correct for \pz{}-based
        tomographic selection bias by propagating the analytical
        \anacal{} flux responses $\partial F_b/\partial\gamma_\alpha$
        through the \pz{} code via a finite-difference step
        (equations~\ref{eq:flux_perturb}--\ref{eq:sel_response}). The
        resulting per-bin selection response
        $\mathcal{R}_{\alpha,\mathrm{sel}}^{(t)}$ is added to the
        denominator of the tomographic shear estimator
        (equation~\ref{eq:tomo_shear_estimator}), correcting for
        shear-driven changes in bin membership without requiring
        external image simulations.

    \item As a first sanity check on real data, we validate \pz{}
        estimates derived from \anacal{} fluxes on the DP1
        \texttt{ECDFS} field, the only DP1 field that combines full
        six-band ($ugrizy$) coverage with an extensive spectroscopic
        reference catalogue. Using \FlexZBoost{} on the DP1 test set,
        the six-band \anacal{} configuration reaches
        $\sigmaNMAD \sim 0.03$ and outlier rate $\sim 10\%$ at
        intermediate redshift, and outperforms a four-band ($griz$)
        baseline across the redshift and magnitude ranges, with the
        largest improvement at $z > 1.2$ where the $u$ and $y$ bands
        break colour--redshift degeneracies
        (Figure~\ref{fig:metrics_combined}). Compared with the same
        algorithms run on standard LSST GAaP fluxes \citep{dp1-pz},
        the \anacal{}-flux \pz{} estimates show a smaller bias at high
        redshift for faint galaxies while matching the scatter and
        outlier rate, confirming that the same fluxes can feed both
        the shear and \pz{} estimators without degrading \pz{}
        performance.

    \item We then validate the analytical selection-bias correction on
        LSST-like image simulations at the expected LSST Y10 depth,
        using two \pz{} algorithms accessed through the \RAIL{}
        framework \citep{RAIL}: \FlexZBoost{} (machine-learning) and
        \BPZ{} (template-fitting). Without the correction, the
        multiplicative shear bias from \pz{}-based binning reaches
        $|m| \sim 1\%$--$2\%$ in the high-redshift bins. After
        applying the \anacal{} correction, $|m|$ falls within the LSST
        DESC SRD Y10 requirement $|m| < 3\times10^{-3}$
        \citep{LSSTRequirement2018} for all five tomographic bins and
        both algorithms (Figure~\ref{fig:bias_6bands}), with additive
        biases statistically consistent with zero. The residual after
        correction is larger for \BPZ{} than for \FlexZBoost{},
        consistent with the larger scatter and outlier rate of the
        template-fitting algorithm.
\end{enumerate}

Together, these results show that \anacal{} provides a self-consistent,
simulation-free pipeline in which the same flux measurements feed both
the shear estimator and the \pz{} estimator, and the \pz{}-induced
selection bias is corrected analytically from the same data.
We caution, however, that the correction requires careful
testing of the full \pz{} pipeline before it is applied to data: we rely on
validation with image simulations for each \pz{} estimation code, as
demonstrated here for \FlexZBoost{} and \BPZ{}. The image simulations serve
only to validate the pipeline; the shear calibration itself does not depend
on them.
Looking ahead, several extensions are planned:

\begin{itemize}
    \item \textbf{Physics-informed AI for shape and flux measurement}:
        Extending the analytical \anacal{} shear-response chain to
        physics-informed AI estimators of both galaxy shape and
        flux. Equivariant neural shape estimators trained on top of
        the \anacal{} baseline have been shown to further reduce
        per-galaxy shape noise \citep{Lin2026_D4CNN_AnaCal}, while
        deep-learning detection,
        deblending, and photometry pipelines such as DeepDISC
        \citep{Merz2023_DeepDISC, Merz2025_DeepDISC_Photoz} deliver
        more accurate multi-band fluxes in crowded fields. Because
        the analytical pixel-level shear response is differentiable,
        it can be propagated through these neural networks to yield
        analytical shear responses of the AI-derived shapes and
        colours; these responses feed both the shear and \pz{}
        estimators within a single self-consistent calibration
        framework, tightening the precision of LSST tomographic
        weak-lensing cosmology.

    \item \textbf{SOM-based tomographic binning}: The Dark Energy
        Survey Year~6 cosmic-shear analysis assigns source galaxies
        to tomographic bins via the SOMPZ framework, in which a
        self-organising map is trained on the deep-field
        multi-band photometry and each wide-field source is binned by
        the SOM cell it falls into rather than by a point estimate of
        its redshift \citep{Yin2025_DESY6_SOMPz}. SOMPZ binning is
        beyond the scope of this work; in future work we plan to
        enable SOMPZ-based tomographic binning within the \anacal{}
        framework by propagating the analytical \anacal{} flux
        responses through the SOM assignment, so that the
        corresponding selection-bias correction follows from the
        same self-consistent chain rule used here.

    \item \textbf{Cross-redshift blending}: When the components of a
        blend lie at substantially different redshifts, each
        component experiences a different shear, and the effective
        redshift distribution probed by the lensing measurement
        departs from the redshift distribution estimated from the
        source catalogue. This mismatch enters as a bias in the
        effective redshift distribution---of order one percent at
        LSST depth---which traditionally must be calibrated with
        external image simulations
        \citep{DESY3_BlendshearCalib_MacCrann2021, Zhang2026_zmixing}.
        Within the \anacal{} framework, two complementary mitigation
        strategies are being pursued: catalog-level removal of
        cross-redshift blends with machine-learning classifiers
        trained on overlapping space-based imaging, which directly
        reduces this mismatch at the sample level
        \citep{Adari2026_CrossZBlend}; and forward modelling of the
        residual effective redshift distribution with controlled
        image simulations along the lines of \citet{Zhang2026_zmixing}.
        Both extensions lie beyond the analytical, simulation-free
        correction presented here.

    \item \textbf{Euclid and Roman synergies}: \pz{} performance in
        the LSST high-redshift regime is limited by colour--redshift
        degeneracies between the $4000~\text{\AA}$ break and the
        Lyman break that are difficult to resolve with optical
        photometry alone. Adding near-infrared photometry from Euclid
        \citep{Euclid2011} or Roman \citep{Roman2020} to the LSST
        $ugrizy$ baseline is forecast to reduce the catastrophic
        \pz{} outlier rate by $\sim 25$--$40\%$ at $z \gtrsim 1$
        \citep{Graham2020_PhotoZ_NIR, Jain2015_LSST_Euclid_WFIRST}.
        \citet{dp1-pz} have already demonstrated such gains on DP1
        \texttt{ECDFS} when Euclid NIR bands are included alongside
        LSST optical photometry. Because the
        \anacal{} selection-bias correction acts on the
        \pz{}-induced selection rather than on the absolute
        \pz{} accuracy, smaller \pz{} scatter and a lower outlier
        rate at high redshift directly reduce the uncorrected
        tomographic shear bias and tighten the residual after
        correction. Joint pixel-level processing of LSST with Euclid
        and, in the longer term, Roman is therefore a natural
        extension of the analytical pipeline presented here.
\end{itemize}

In summary, this work demonstrates that the \anacal{} analytical
calibration extends naturally to photometric-redshift-based tomographic
selection, providing a practical and accurate calibration pathway for
the LSST and other Stage~IV weak lensing surveys.

\section*{ACKNOWLEDGEMENTS}
\addcontentsline{toc}{section}{ACKNOWLEDGEMENTS}

We thank the \RAIL{} team and Erin Sheldon for useful discussions, and
the LSST DESC Weak Lensing and Photo-$z$ Working Groups for feedback
throughout this work.

This paper makes use of LSST Science Pipelines software developed by
the Vera C.\ Rubin Observatory. We thank the Rubin Observatory for making
their code available as free software at \url{https://pipelines.lsst.io}.
This work also makes use of the
open-source scientific Python ecosystem, including \numpy{}
\citep{numpy_Harris2020}, \texttt{SciPy} \citep{scipy_Virtanen2020},
\texttt{Matplotlib} \citep{matplotlib_Hunter2007}, \galsim{}
\citep{GalSim}, and the \texttt{conda-forge} distribution
\citep{conda_forge}.

Xiangchong Li acknowledges support from the U.S.\ Department of Energy (DOE)
under Contract No.\ DE-SC0012704 and from the Laboratory Directed Research and
Development Program at Brookhaven National Laboratory (Project No.\ 27992). TZ
is supported by Schmidt Sciences.  RM is supported in part by the Department of
Energy Cosmic Frontier program, grant DE-SC0010118. This paper has undergone
DESC internal review. The authors would like to thank the internal reviewers
Robert Armstrong and Erin Sheldon.

The DESC acknowledges ongoing support from the Institut National de
Physique Nucl\'eaire et de Physique des Particules in France; the
Science \& Technology Facilities Council in the United Kingdom; and the
Department of Energy and the LSST Discovery Alliance in the United
States. DESC uses resources of the IN2P3 Computing Center
(CC-IN2P3--Lyon/Villeurbanne - France) funded by the Centre National de
la Recherche Scientifique; the National Energy Research Scientific
Computing Center, a DOE Office of Science User Facility supported by
the Office of Science of the U.S.\ Department of Energy under Contract
No.\ DE-AC02-05CH11231; STFC DiRAC HPC Facilities, funded by UK BEIS
National E-infrastructure capital grants; and the UK particle physics
grid, supported by the GridPP Collaboration. This work was performed
in part under DOE Contract DE-AC02-76SF00515.

Author contributions to this work are as follows: Xiangchong Li developed
the method for correcting the selection bias due to tomographic binning,
developed the pipeline for the correction, and wrote the paper. Tianqing
Zhang provided the photometric redshift estimation and prepared the
training sample in LSST DP1. Rachel Mandelbaum supervised the project,
providing valuable discussions and feedback on the paper.

\section*{DATA AVAILABILITY}
\addcontentsline{toc}{section}{DATA AVAILABILITY}

The \anacal{} code is publicly available on GitHub:
\url{https://github.com/mr-superonion/AnaCal}.
The DP1 photo-$z$ data products are publicly available through the
Rubin Science Platform (\url{data.lsst.cloud}) and the Photo-$z$ Server
(\url{https://pzserver.linea.org.br}).

\section*{AI USAGE DISCLOSURE}
\addcontentsline{toc}{section}{AI USAGE DISCLOSURE}

Artificial intelligence (AI) tools have been used to assist with
improving the English writing of this paper. All scientific
content, analysis, and conclusions are the authors' own, and the
authors have reviewed and take full responsibility for the final
text.

\bibliographystyle{aasjournal}
\bibliography{citation}

\end{document}